\documentclass[letterpaper, 10 pt, conference]{ieeeconf} 
\usepackage{cite}
  \usepackage[pdftex]{graphicx}
  \graphicspath{ {./images/} }
\DeclareGraphicsExtensions{.pdf,.jpeg,.png}

\usepackage{amsmath}
\usepackage{amssymb}

\usepackage{amsthm}

\usepackage{algpseudocode}
\usepackage{algorithm}

\newtheorem{definition}{Definition}
\usepackage{xcolor}
\usepackage{hyperref}
\usepackage{caption}
\usepackage{subcaption}
\usepackage[affil-it]{authblk}
\usepackage{diagbox}
\begin{document}
%\title{Do you see me? T-PoP: A Decentralised and Collaborative, Tree-based Proof of Position Algorithm}

\title{Robust decentralised proof-of-position algorithms for smart city applications}
\author[1]{Aida Manzano Kharman  \texttt{amm3117@ic.ac.uk}}
\author[1]{Pietro Ferraro}
\author[1,2]{Anthony Quinn}
\author[1]{Robert Shorten}
\affil[1]{Imperial College London, Dyson School of Design Engineering}
\affil[2]{Trinity College Dublin, Electronic and Electrical Engineering}

\maketitle

\begin{abstract}

We present a decentralised class of algorithms called Tree-Proof-of-Position (T-PoP). T-PoP algorithms rely on the web of interconnected devices in a smart city to establish how likely it is that an agent is in the position they claim to be. T-PoP operates under adversarial assumptions, by which some agents are incentivised to be dishonest. We present a theoretical formulation for T-PoP and its security properties, and we validate this model through a large number of Monte-Carlo simulations. We specifically focus on two instances of T-PoP and analyse their security and reliability properties under a range of adversarial conditions. Use-cases and applications are discussed towards the end of this paper.   
\end{abstract}

\newcommand{\tpop}{\textsf{T-PoP}}

% \fi
\IEEEpeerreviewmaketitle

\section{Introduction}
A basic problem across a range of application areas is the need for decentralised agents to be able to certify their position in a trustworthy and certifiable manner. For example, in crowd-sourcing applications arising in the context of smart cities, the need for agents to certify their position in a trustworthy manner is essential; one such use-case arises when vehicle cameras are used to identify parking spot locations or vacant or available electric charge points \cite{cogill2014parked}. Other examples of this nature are emerging in the context Smart Mobility applications when vehicles need to prove their location to avail of certain services; for example, in the case of hybrid vehicles using their electric engine mode in a city to avoid an environmental charge (as in London); when making use of a fast or slow lane on a highway and paying the associated charge; or when infotainment services are offered to vehicles when adopting certain positions.

Our objective in this paper is to propose a suite of algorithms whereby agents may certify their position collaboratively, but in a decentralised manner. Our algorithms are designed to be robust in the sense that they do not require the use of centralised infrastructure, and in the sense that they are designed to operate successfully in an adversarial environment (in the presence of agents that are interested in coercing the system for their own personal objectives). The need to be independent of a centralised authority is fundamental to our work, as such authorities may be compromised or a subject to data and privacy leaks \cite{fiesler2018we}. While our original motivation arises from automotive applications, the work presented here is relevant and may find application in other disciplines and applications, and may also help to encode basic elements of fairness, social justice and civil rights.  More specifically, in an era characterised by fake news, and deep fake technology, the ability to associate sensing information with a verifiable geographic position, is not only essential in establishing the veracity of sensed information, but also in developing robust decision making analytics based on this data. Currently across many such applications, sensed information is assumed more trustworthy if a number of people agree on it. In scenarios where we cannot verify ourselves what happened, we search for `truth' by listening to our peers and believing what a majority claims \cite{desai2022getting}. 
%Whilst this may carry its risks, it is often how our society operates, whether this be in democracy, peer reviewing research \cite{chinn2018consensus} or accepting news that are consistent across outlets. 
%However, in an increasingly digital age, fact checking becomes hard and misinformation, rife. We note that for a wide class of social applications, claims can be made of an event happening at a given location and time. Having a way to prove that you were indeed where you claimed to be, is a first step in building trust about such statements. Then, if a large number of agents agree with your statement, and were also near you at that time, that statement becomes more trustworthy. 
So our research question becomes: how can we provide agents with the ability to claim that they are at a given place in time, without hinging the security of our protocol on the honesty of a centralised power? While we are not the first to attempt to address the aforementioned research question, upon exploring existing solutions, we found that none addressed the requirements of applications in smart city contexts. Namely, the solution must be truly decentralised, and it must be robust to attacks whilst preserving user privacy.

Our work is motivated by recent developments in distributed ledger technologies (DLT); in particular, in the design of distributed acyclic graph based distributed ledgers. However, while the design of such ledgers is concerned with architectures that can provide peer-to-peer trustworthy record keeping, we are interested in realising DAG-based algorithms that encode reliable position information.

\subsection{Related Work} 
Several papers been published on the topic of proof-of-position; see for example \cite{nosouhi2018sparse}, \cite{alamleh2020cheat}, \cite{javali2016alice}, \cite{wu2020blockchain}, \cite{amoretti2018blockchain}. Most of these papers are unsuitable for the type of applications that we are interested in due to unrealistic trust assumptions and \textit{de facto} centralisation in the systems that they propose. In the remainder of this section we give a snapshot of some of this prior work. 

An early example of a decentralised proof-of-location scheme, termed APPLAUS, was presented in \cite{zhu2011toward}. The APPLAUS scheme makes a number of valuable contributions; namely it looks to address collusion attacks using graph clustering and computing a `betweeness' metric. In \cite{zhu2011applaus}, nodes in the graph that are weakly connected are considered less trustworthy. They also present a weight function that decays with time, and compute the trustworthiness of a node by calculating a node's ratio of approvals to neighbours. These contributions serve as a starting point to the work here presented. However, in their work, users must register their public/private keys with a trusted Certificate Authority, thereby breaking an assumption of being truely decentralised. A focal point of our work is that we do not assume a trusted centralized authority, and indeed we argue that introducing this assumption makes a system de-facto centralised and poses security and privacy risks. Another algorithm known as SHARP is introduced in  \cite{zheng2012sharp}. Here, the authors present a private proximity test that does not require a user to reveal their actual location to a server, and furthermore, they present a secure handshake method wherein users do not need to have a pre-shared secret. A noteable contribution in this work is that a witness\footnote{An agent that will verify to seeing another agent wishing to prover their position.} may only extract the session key if they are indeed in the vicinity of the prover \footnote{An agent that wishes to prove their position.}. The security metric in this work is to ensure that the location tags are unforgeable, thus implying that the protocol is robust towards location cheating. A weakness of the protocol is that a user in a given location can generate a valid proof and they could then relay this valid proof to a malicious agent that is not in the same location as them. Another algorithm known as Vouch + is presented in \cite{boeira2019decentralized}. This is another decentralised approach to prove location, with a focus on addressing high speed and platooning scenarios. The major disadvantage of the work presented is that its security relies on selecting a proof provider that is honest. This assumption, in our opinion, is too strong. We aim to develop a protocol wherein the prover could lie, and the system would still have a probabilistic guarantee of detecting this. As another example, SPARSE, the protocol presented in \cite{nosouhi2018sparse} does not allow the prover to pick their own witnesses, making collusion significantly harder. Furthermore, SPARSE does address necessary security concerns, and achieves integrity, unforgeability and very importantly: non-transferability. However, similarly to \cite{boeira2019decentralized}, the prover is assumed to be a trusted entity which supposedly does not publish users' identity and data.

\subsection{Contributions}

We present a generalised model for a class of decentralised, proof-of-position algorithms. We also present a mathematical model to describe the operation of this class of algorithm and to facilitate further analysis. Simulations are also presented that validate our mathematical model, and we present a framework for users to tailor the operating conditions of the algorithm to satisfy their security and reliability requirements.
We also provide probabilistic guarantees of detecting dishonest provers and collusion attacks.

\textbf{Comment:}
The algorithm can also be implemented in a privacy preserving manner given that T-PoP does not require the agent running the algorithm to actually reveal their true position, but rather a cryptographic commitment \cite{oded2001foundations} to one's position suffices. Depending on the security requirements of the application, T-PoP users can pick a commitment scheme with varying binding and hiding, as long as the commitment scheme supports the computation of Euclidean distance between two points. 

Finally, we do not constrain the freedom of adversarial agents to misbehave. We consider not only the possibility of them being dishonest about their own position, but also colluding to lie about other agents' position.

\subsection{Structure of the paper}
Our paper is structured as follows: first we introduce the T-PoP protocol and explain its functioning in section \ref{TPOP section}, next we present a theoretical model for the T-PoP class of algorithms in section \ref{theoretical section} and finally we simulate T-PoP in a more realistic scenario in section \ref{sec:sim}, thus validating our theoretical model too. 

\section{Tree - Proof of Position protocol} \label{TPOP section}
\begin{figure}
    \centering
    \includegraphics[scale = 0.19]{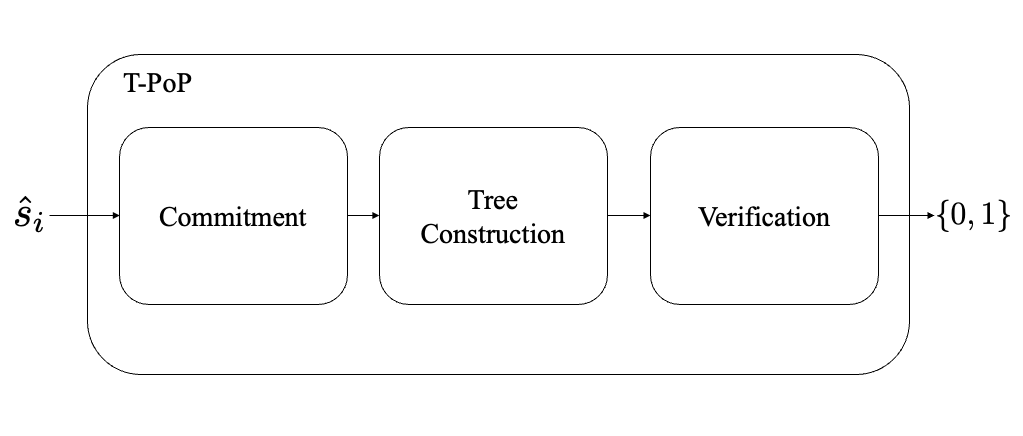}
    \caption{High-level Overview of the T-PoP protocol}
    \label{fig: TPoPArch}
\end{figure}

\begin{figure}
    \centering
    \includegraphics[scale = 0.23]{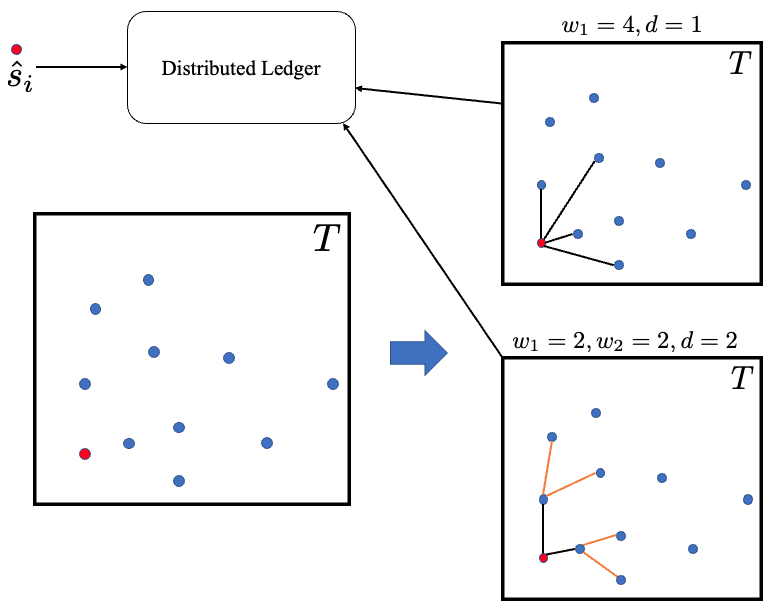}
    \caption{Tree building examples.  Agent $a_i$ commits their alleged position $\hat{s}_i$ to a distributed ledger. The panel on the top right shows the construction of a tree for $d = 1$ and $w_1 = 4$, while the panel on the bottom right shows the construction of a tree for $d = 2, w_1 = 2, w_2 =2$.}
    \label{fig: T-PoPTree)}
\end{figure}

We begin by providing a high level explanation of how the protocol operates. Subsequently, we will provide the necessary definitions for each stage of  and explain them in a detailed manner. We assume that agents willing to participate in the protocol are situated in a two dimensional area $T \subseteq \mathbb{R}^2$ (the protocol can be seamlessly extended to a three-dimensional space). Each agent $a_i$  is characterised by their \emph{true} position  $s_i = (x_i, y_i) \in T$ and by their \emph{claimed} position $\hat{s}_i = (\hat{x}_i, \hat{y}_i) \in T$ while the set of all agents is denoted by $A$. Notice that it is possible that $\hat{s}_i \neq s_i$ (in the event an agent is lying). An agent, $a_j$, is (allegedly) $a_i$'s  neighbour if $||\hat{s}_i - \hat{s}_j|| < r_i$, where $r_i > 0$ is each agent's range-of-sight. T-PoP is performed in three steps, as depicted in Figure \ref{fig: TPoPArch}:
\begin{itemize}
    \item \emph{\bf Commit}: At the beginning of T-PoP, each agent, $a_i \in A$, commits to their claimed position, $\hat{s}_i$ nd publishes $\hat{s}_i$ on a distributed ledger (DL). This ensures that the agent's commitment\footnote{The only necessary requirement for our protocol is that the commitment is binding \cite{oded2001foundations} To ensure user privacy, we favour schemes that allow for the computation of the Euclidean distance between two points which  can be achieved by leveraging encryption schemes that are fully homomorphic. It is also necessary to achieve non-repudiation, which can be done through the use of digital signatures. Frequently used examples include  \cite{bernstein2011high} and \cite{johnson2001elliptic}.  This ensures an agent cannot later deny having claimed to be in a given position \cite{swanson2006guide}. Finally, non-transferability is needed to ensure that if an honest prover generated a valid location proof through \tpop, they cannot then transfer their honest proof to a malicious actor. A user's identity is unique upon being issued, and should this be in the form of a private key, we introduce the assumption that users do not share it.} cannot be changed later.

    \item \emph{\bf Tree Construction}:  Each agent, $a_i$, then constructs a tree of depth $d \in \mathbb{N^+}$, incorporating the committed positions of agents, called \emph{witnesses}, at levels $l\in\{0,\ldots,d\}$. A specific $a_i$---which we denote as $g$---is the root of the resulting tree.  
    These $g\in A$-indexed trees are also committed to the DL as they are part of the proof-of-position protocol. For every \emph{prover}, $g$, the tree is constructed as follows: 
     \begin{itemize}
         \item $g$ is is the root node at level 0. 
         \item For each $l \in \{1,...,d\}$, each node at level $l-1$ will name $w_{l}$ \emph{witnesses}. A witness at level $l$ is an agent, $a_j$, that is a neighbour (see above) of a witness, $a_i \in W_{l-1}$, at level $l-1$ (note that if $\hat{s}_i \neq s_i$ and $a_i$ is lying about their position it is possible that $a_i$ and $a_j$ might not actually be \emph{true} neighbours). $a_i$ is called the $parent$ of witness $a_j$. The set of all witnesses at level $l$ is called $W_l$, with $|W_l|\equiv n_l$. 
         \item If  $a_j$, was named a witness at some point in the tree, it should not be named again by another agent. Otherwise, if this happens, the prover will be considered dishonest.
     \end{itemize}
     In practice, the root node $g$, names $w_1$ witnesses who in turn would each name $w_2$ witnesses and so on, until we reach depth $d$.  The number of witnesses per level, $n_l$, can therefore be computed recursively:

     \begin{equation}
         n_l = w_l n_{l-1},\;l=1,\ldots,d,
         \label{eq:nl}
     \end{equation}
    with $n_0 \equiv 1$. Figure \ref{fig: T-PoPTree)} depicts the operation of this process.

    \item \emph{\bf Verification}: The agent wishing to prove their position runs the verification stage with the tree as an input, initialized with $l = d$.
    \begin{itemize}
        \item[1] Each  witness at level $l$ states whether their parent at level $l-1$ is their neighbour or not. If the answer is yes, and the witness has not yet been named in the tree, this witness becomes a confirmed level $l$ witness. The total number of confirmed level $l$ witnesses is denoted as $M_l \leq n_l$, and the total number of witnesses that confirm parent $b$ at any level, $l$, is denoted by $K_b \leq w_l$. It follows that
        \begin{equation}
            M_l = \sum_{b \in W_l} K_b \leq n_l
        \end{equation}
        %and so the total number of confirmed witnesses of $a_i$ is $M\equiv \sum_{l=1}^d M_l$. \textcolor{red}{AQ: problematical that $M$ not indexed by $i$ but it's alright for now.}
        \item[2] If $K_b<t \cdot w_l$,  $t \in (0,1]$, parent $b$ is eliminated from the tree. Here, $t$ is a parameter of T-PoP, called the {\em threshold}, which is used to regulate the Security and Reliability properties of the algorithm, defined in Section III.
        \item[3] If $M_l <  t \cdot n_l$ then the algorithm interrupts and outputs that root $g$ is lying about their position. Otherwise we move on to level $l-1$ and we repeat this process. Note that any parent removed by the previous step will not be included in this next iteration of T-PoP.
    \end{itemize}
\end{itemize}

T-PoP is therefore an algorithm depending on  a set of parameters, $\theta \equiv \{t, d, w_1, ..., w_d\}$.   The influence of these parameters on the performance of the algorithm will be explored in Section IV,  via two examples. The pseudo-code for the \emph{Tree Construction} and \emph{Verification} stages of the protocol can be found in Algorithms \ref{tree algorithm} and \ref{verification algorithm} respectively. 

\begin{figure}[h]
     \centering
     \begin{subfigure}[t]{0.46\textwidth}
         \centering
         \includegraphics[width=0.49\textwidth]{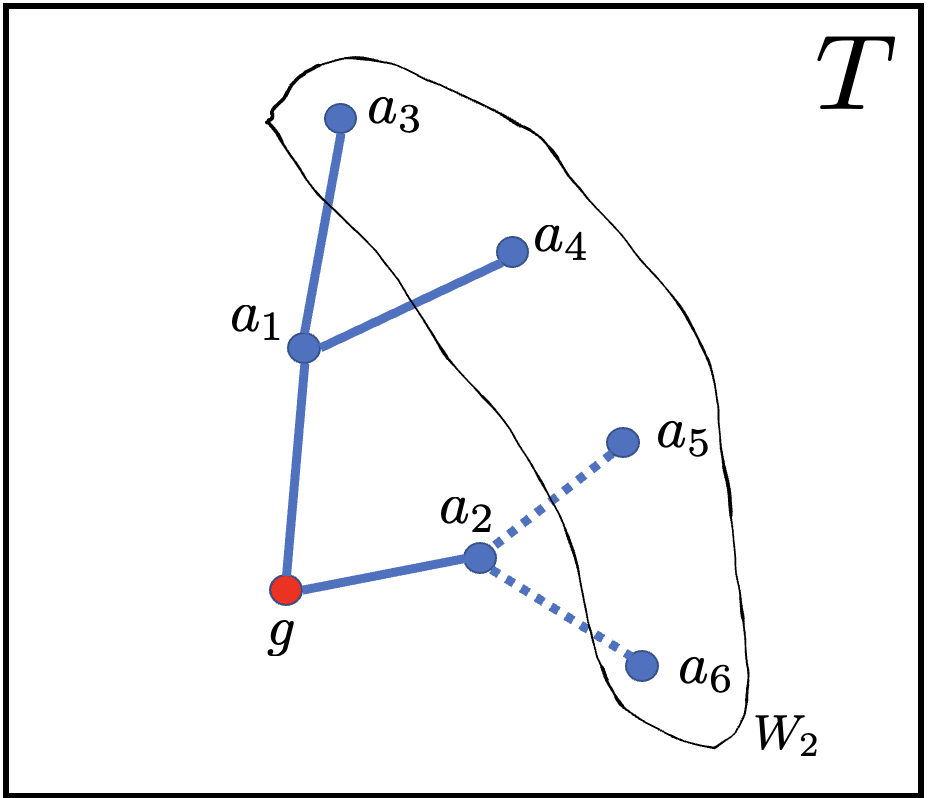}
         \caption{We start by evaluating the outer level of the tree and we evaluate the witnesses in $W_2$. Agents $a_5$ and $a_6$ while $a_2$ is a honest agent. Therefore $a_5$ and $a_6$ will not confirm that they see agent $a_2$. This leads to agent $a_2$ being eliminated from the tree.}
         \label{fig:example a}
     \end{subfigure}
     \hfill
     \begin{subfigure}[t]{0.46\textwidth}
         \centering
         \includegraphics[width=0.49\textwidth]{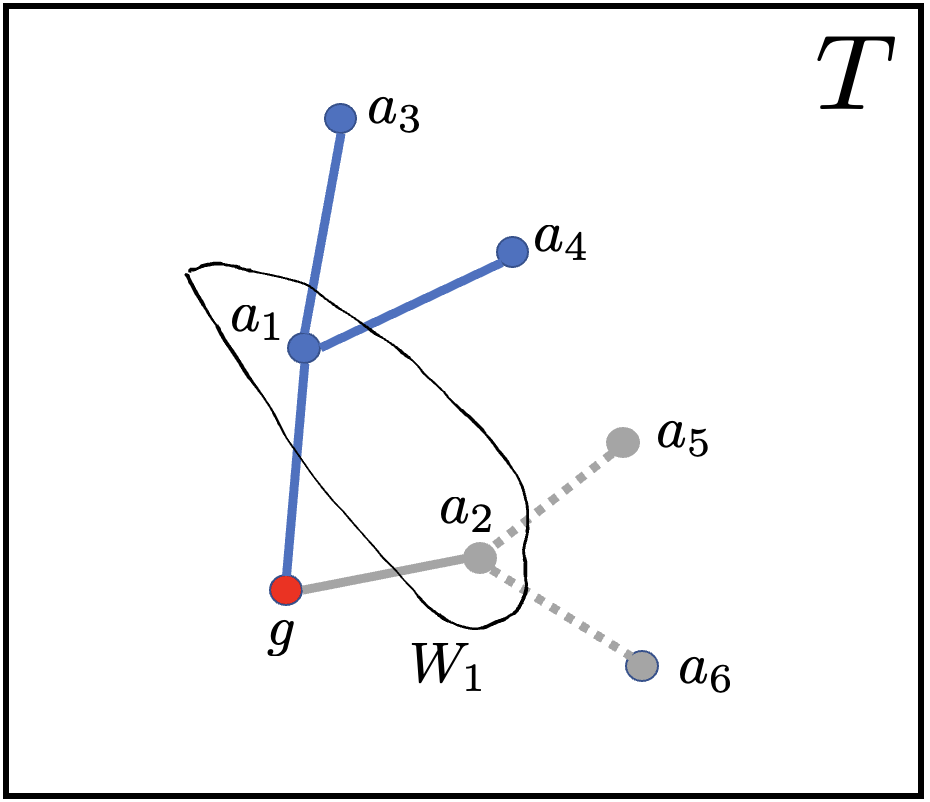}
         \caption{We go down one level and now we evaluate the witnesses in $W_1$. $a_2$ has been eliminated by the tree (shown in grey in this image) and all the is left is $a_1$.}
         \label{fig:example b}
     \end{subfigure}
        \caption{Example of T-PoP algorithm with $d=2, w_1 = 2, w_2 =2$.}
        \label{fig:example}
\end{figure}

{\bf Example:} Consider the  T-PoP example in Figure \ref{fig:example}, in which  $\theta = \{t =0.5, d = 2, w_1 = 2, w_2 = 2\}$, and so $n_1 = 2$ and $n_2 = 4$ (\ref{eq:nl}). Solid arrows mean that a witness approves their parent and dotted lines mean that a witness does not approve their parent. Agents $a_5$ and $a_6$ are  dishonest agents, so that their committed positions, $\hat{s}_5$ and $\hat{s}_6$, are different from their true positions. However, agent $a_2$  does not know this, it saw those cars next to it and it picked $a_5$ and $a_6$ as witnesses. So, $a_5$ and $a_6$ do not confirm that $a_2$ is a neighbour of theirs, whereas $a_3$ and $a_4$ confirm that $a_1$ is a neighbour of theirs. In line with point 2 of \emph{Verification} (above), agent $a_1$ has enough confirmed witnesses ($K_{a_1} = 2 \geq t\times w_2 =  0.5 \times 2$) and stays in the tree, while agent $a_2$ does not have enough confirmed witnesses ($K_{a_2} = 0 < 0.5 \times 2$), and so $a_2$ is removed from the tree. However, since the total number of confirmed witnesses at level 2 is $M_2 = 2 \geq t \times n_2  = 0.5 \times 4 $, T-PoP does not stop for $g$ (\emph{Verification}, point 3), and we move to level 1. At level 1, $a_2$ has been removed but $a_1$ confirms that $g$ is its neighbour. As per points 2 and 3 of \emph{Verification}, the final output of T-PoP is that $g$ is {\em truthful\/} about their position. $t$ is critical in determining the output of T-PoP.  For instance, if $t=1$, then $M_2 = 2 < t \times n_2  = 1 \times 4 = 4$, causing T-PoP to stop at point 3 of \emph{Verification}, and returning an output of {\em untruthful\/}  for $g$. 

\subsection{Possible Adversarial Behaviours}

In order to analyse the properties of T-PoP, we introduce two  qualities that each agent, $a_i \in A$, will exhibit:
\begin{definition}[Honest and Dishonest agents]
Every $a_i\in A$ is either \emph{honest} or \emph{dishonest}.
The set of honest agents is denoted by $H \subseteq A$, and the set of dishonest agents is denoted by $\overline{H}$.
A dishonest agent will always commit a position $\hat{s}_i \neq s_i$. A honest agent on the other hand will always commit a position $\hat{s}_i = s_i$.
\end{definition}
\begin{definition}[Coerced and Non-Coerced Agents]
Every $a_i\in A$ is either  \emph{coerced} or \emph{non-coerced}. 
The set of coerced agents is denoted by $C \subseteq A$, and the set of non-coerced agents by $\overline{C}$.
A coerced agent will claim to see agents that are not actually in its vicinity, if the latter are dishonest. 
\end{definition}
$a_i$ will interact  with its neighbours in different ways---as defined next---depending on which of the four possible states it falls into with respect to the two 2-state classes above.
\begin{definition}[Neighbour-adding logic]
    Every agent, $a_i \in A$,  adds neighbours, $a_j$, according to the following logic:
    \begin{itemize}
    \label{neighbours adding}
    \item If $a_i\in 
    \overline{H}$, it can add $a_j$ as a neighbour if $a_j$'s position, is within the range-of-sight $r_i$, of $a_i$'s fake position, $\hat{s}_i \neq s_i$. This implies that $a_i$ checks who is in the $r_i$-neighbour of the fake position that they committed.
    \item If $a_i\in H$, it can add $a_j$ as a neighbour if $a_j$'s committed position is within the range-of-sight, $r_i$, of $a_i$'s true position, $s_i$.
    \item If $a_i \in \overline{C}$, it can only add $a_j$'s true position, $s_j$, if this is within $a_i$'s range-of-sight, $r_i$.
    \item If $a_i \in C$, it can add $a_j$'s true position, $s_j$, if $a_j$ is honest, and its fake position, $\hat{s}_j$, if $a_j$ is dishonest. 
\end{itemize}
\end{definition}

%Different amount of honest or dishonest and coerced or non-coerced agents greatly affect the outcome of T-PoP, as we will investigate in Section~\ref{sec:sim}.

%Once this stage is completed, we can proceed with an agent claiming to be in a given position. They commit to said position using a cryptographic commitment scheme. The only necessary requirement for our protocol is that the commitment is binding. Cryptographic commitment schemes are computationally or perfectly hiding and binding \cite{oded2001foundations}. To ensure a user's privacy, we favour schemes that allow for the computation of the Euclidean distance between two points, whilst these points are still encrypted. This can be achieved leveraging encryption schemes that are fully homomorphic. It is also necessary to achieve non-repudiation, which can be done through the use of digital signatures. Frequently used examples include: \cite{johnson2001elliptic} and \cite{bernstein2011high}. This ensures an agent cannot later deny having claimed to be in a given position \cite{swanson2006guide}. Finally, non-transferability is needed to ensure that if an honest prover generated a valid location proof through \tpop, they cannot then transfer their honest proof to a malicious actor. A user's identity is unique upon being issued, and should this be in the form of a private key, we introduce the assumption that users do not share it. 

\begin{algorithm}
\caption{Tree Construction} 
	\begin{algorithmic}[1]
            \Require{Prover $a_i$, Depth $d$, Number of witnesses $w_1, ..., w_l$}
            \State{Initialise $a_i$ as the root of the tree $G$ and as a witness of level $0$}
                \For{$l = 0, 1, ..., d - 1$}
                    \For{Each witness $a$ at level $l$}{}{}
                        \State
                        \parbox[t]{\dimexpr\linewidth-\algorithmicindent}{$a$ names $w_{l+1}$ witnesses among its neighbours}
                        \State
                        \parbox[t]{\dimexpr0.91\linewidth-\algorithmicindent}{All the named neighbours are added as nodes of $G$ at level $l+1$, with $a$ as parent node}
                    \EndFor
                \EndFor
            \State{{\bf return} G}
	\end{algorithmic} 
        \label{tree algorithm}
\end{algorithm}

\begin{algorithm}
\caption{Verification} 
	\begin{algorithmic}[1]
            \Require{Tree $G$, Threshold $t$}
            \State Initialise $M_0, M_1, ..., M_{l-1}$ to $0$
            \For{$l = d - 1, d - 2, ..., 0$}{}{}
                \For{Each witness $a$ at level $l$}{}{}
                    \State Set $C = 0$
                    \For{{Each $b$ that has been named by $a$}}
                        \If{$b$ confirms $a$ and $b$ unique in $G$}
                            \State $C \longleftarrow C + 1$
                            \State $M_l \longleftarrow M_l + 1$
                        \EndIf
                    \EndFor
                    \If{ $C<t w_{l+1}$}
                        \State Remove $b$ from $G$
                    \EndIf
                \EndFor
                \If{$M_l<\#\{\text{witnesses at level $l + 1$}\} t$}
                    \State{{\bf return} False}
                \EndIf
                \EndFor
            \State{{\bf return} True}       

	\end{algorithmic} 
        \label{verification algorithm}
\end{algorithm}

%\emph{Comments:}

%Other elements that we will take into account, for when we try to look at the dynamic of the system are: the consistency of the position of agent $i$ over multiple PoLs and the entropy of the range-of-sight function (i.e., if always the same cars validate you, while you are moving, there might be something fishy).

\section{Theoretical Analysis} \label{theoretical section}
\begin{figure}[h]
    \centering
    \includegraphics[width=0.75\columnwidth]{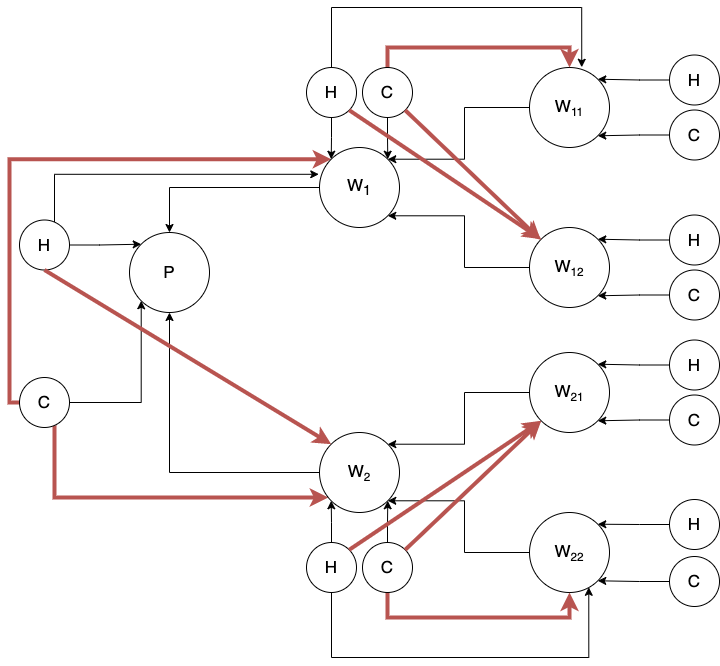}
    \caption{Probability Model of T-PoP with parameters $d=2, w_1=2, w_2=2$. The red lines indicate that those variables influence the output of a specific node.}
    \label{fig:Probability Model}
\end{figure}

The stochastic nature of T-PoP is modelled via the probabilistic graphical model  in Figure \ref{fig:Probability Model},  for the  case where  $d=2, w_1=2, w_2=2$. We assume that the Honesty and Coercion states of each agent are independently and identically distributed (iid) Bernoulli trials. Formally, for each agent, we define two independent random variables, $h\sim \mathcal{B}(p_h)$ and $c\sim \mathcal{B}(p_c)$, where $p_h\in[0,1]$ and $p_c\in[0,1]$ are the probabilities of any agent being honest and coerced, respectively (and it follows that $1-p_h$ and $1-p_c$ are the  probabilities of an agent being respectively dishonest and non-coerced). Depending on the outcome of these trials for a witness at level $l$, it will then deterministically confirm that the witness at level $l-1$, which named them, is its neighbour or not (note that agents might be lying about whether another agent is their \emph{true} neighbour or not). The outcome of this interaction has been described in definition \ref{neighbours adding}, and is summarized in the truth table (Table \ref{tab:truth table}). If agent, $a_i$, verifies agent $a_j$'s position, the outcome is 1, and 0 otherwise. 
\begin{table}[h]
    \centering
    \def\arraystretch{1.5}
    \begin{tabular}{|c|c|c|c|c|}
        \hline
         \diagbox[width=5em]{$a_i$}{$a_j$} & $h$ and $\overline{c}$ & $h$ and $c$ & $\overline{h}$ and $c$ & $\overline{h}$ and $\overline{c}$ \\
        \hline
        $h$ and $\overline{c}$  & 1 & 1 & 1 & 1 \\
        \hline
        $h$ and $c$ & 1 & 1 & 0 & 0 \\
        \hline
        $\overline{h}$ and $c$ & 1 & 0 & 1 &  0 \\
        \hline
        $\overline{h}$ and $\overline{c}$  & 1 & 0 & 0 & 1 \\
        \hline
         
    \end{tabular} 
    \caption{A truth table showing confirmation (1) or rejection (0) of a parent's ($a_i$) position by a witness  ($a_j$), depending on the  honesty ($h$) and coercion ($c$) states of each agent. Notice that the relationship between $a_i$ and $a_j$ is symmetrical.}
    \label{tab:truth table}
\end{table} 
In this model, we assume that the density of agents in $T$ is very high. This means that while provers construct their tree following Algorithm \ref{tree algorithm}, they are always able to find $w_l$ witnesses at each level and that each witness is always unique. While this assumption might sound unrealistic, as in many cases agents might be alone and not have enough witnesses around them, we believe that studying the outcome of the model in this high-density scenario provides a good assessment of the qualities of T-PoP. Indeed, we argue that if an agent is honest but does not have sufficient witnesses, it is fair to consider them less trustworthy. Once the tree has been created, the \emph{Verification} step can be used to provide the outcome of  the algorithm, which can be either 0 (if the algorithm deems the prover dishonest) or 1 (if the algorithm deems the prover honest). Given a prover, $g$ (the root of the tree), we define a random variable, $C(g) \in \{0,1\}$, whose outcome depends on the ensemble of iid random variables, $h, c$, in its constructed tree, and on T-PoP parameters, $\theta \equiv \{t, d, w_1, ..., w_d\}$. In order to analyse T-PoP's performance, we consider two metrics: reliability and security. 

\begin{definition}
    {\em Security, $S$, \/} is a conditional probability  quantifying the ability of the algorithm to detect malicious agents. Specifically, it is the true-negative conditional probability, which, under stationarity assumptions, is independent of $i\in \{1, \ldots,|A|\}$:
    \[
S \equiv \Pr[C(g) = 0 | a_i \in \overline{H} ]
\]
 \end{definition}

\begin{definition}
    {\em Reliability, $R$,\/} is a conditional probability  quantifying the ability for the algorithm to detect honest agents. Specifically, it is the true-positive conditional probability.
    Once again, under stationarity assumptions: 
 \[
R\equiv\Pr[C(g) = 1 | a_i \in H ]
\]
\end{definition}

\begin{figure*}[h]
     \centering
     \begin{subfigure}[t]{0.7\textwidth}
            \centering
            \includegraphics[width=\textwidth]{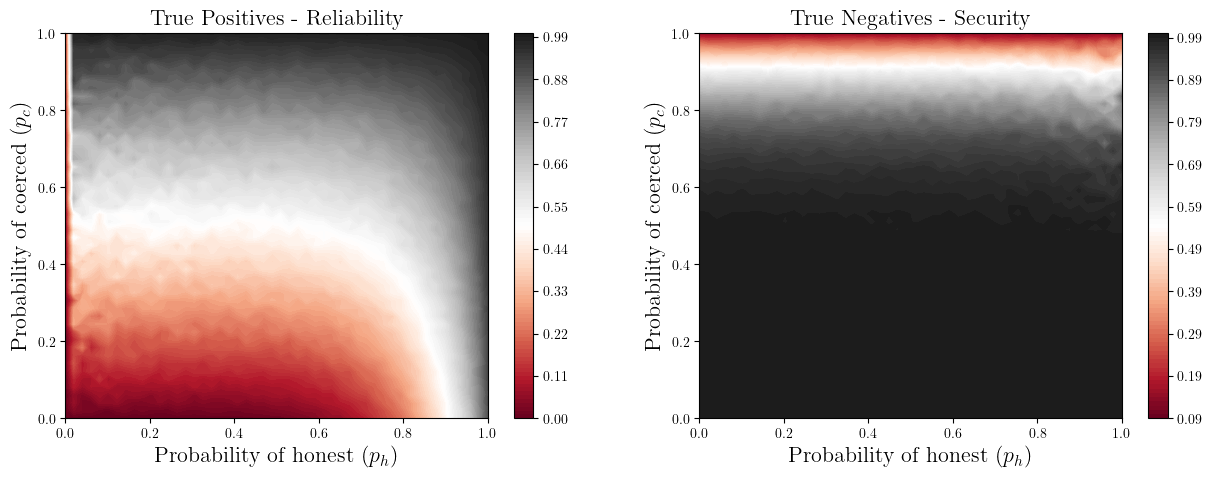}
            
     \end{subfigure}
     \hfill
     \begin{subfigure}[t]{0.7\textwidth}
             \centering
            \includegraphics[width=\textwidth]{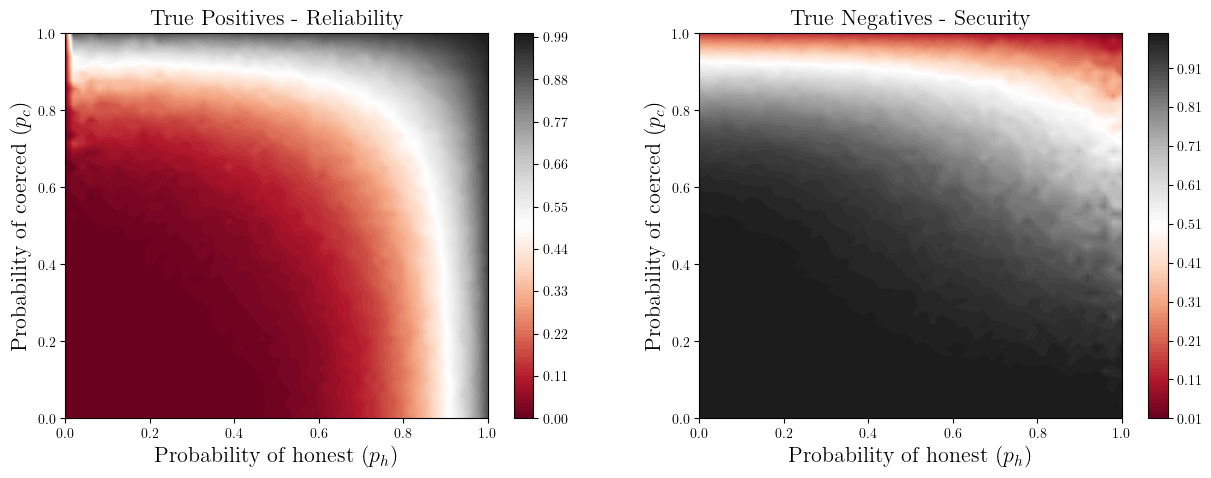}
            
     \end{subfigure}
        \caption{T-PoP performance for the {\bf graphical probability model} (Figure~\ref{fig:Probability Model}). The panels in the left column show reliability, $R$, while the panels in the right column show security, $S$. The first row is associated with model parameters, $\theta_1$, while the second row is associated with model parameters, $\theta_2$. }
\label{tpop model figure}
\end{figure*}

In Figure \ref{tpop model figure}, we display empirically evaluated $R$ and $S$ for two sets of parameters, respectively $\theta_1 = \{t = 1, d = 1, w_1 = 6\}$ and $\theta_2 = \{t = 1, d = 2, w_1 = 2, w_2 = 2\}$, varying $p_h$ and $p_c$ in their ranges, $[0,1]$, with steps of 0.02. To emphasize the functional dependence of these probabilistic performance metrics on the honesty and coercion probabilities of the iid agents, we denote these these metrics by $R(p_h, p_c)$ and $S(p_h, p_c)$.  The values for $R(p_h, p_c)$ and $S(p_h, p_c)$ are obtained through empirical evaluation via extensive Monte Carlo simulations (we simulated 5000 trees for each choice of parameters) of the graphical model.

\section{Simulations}
\label{sec:sim}

 In this section we present an agent-based simulator, coded in Python, to replicate a more realistic scenario for T-PoP and to validate the graphical theoretical model that we presented in the previous Section. Each agent has a number of varying attributes such as their range-of-sight, position, velocity, unique identifier and whether they are honest or dishonest and coerced or not. Depending on the latter variables, each agent will commit to their true position or a fake one, and will add agents to their set of neighbours as outlined in definition \ref{neighbours adding}. We then create an environment with a fixed density of agents in it, and place these randomly and uniformly across the environment. We allow them to move according to their velocity vector, within the bounds of the environment. Each time the agents move, all agents construct a new set of neighbours and discard the previous one. 
 Next, each agent wishing to claim their position runs T-PoP; namely, they run the \emph{Tree Construction} and the \emph{Verification} algorithms. Our simulator can be found in this \hyperlink{https://github.com/V4p1d/Tree-Proof-of-Position.git}{GitHub Repository.}. Preliminary simulations show that the density of agents in the environment vastly affected the performance of T-PoP. This was especially noticeable when the average number of agents per range-of-sight in the environment was lower than the total number of nodes of the tree being constructed, which greatly increased the number of False Negatives, thus making T-PoP unsuitable for low density environments. Other key variables are the threshold, depth and number of witnesses used. A greater threshold increases security, but also reduces reliability. Increasing the number of witnesses increased both security and reliability, however, this may not be a suitable measure for sparser scenarios, or cases where agents are moving at high speed, and may cause a communication overhead. We advocate for the users to select the appropriate threshold, depth and number of witnesses based on the individual needs of their own application. Lowering the threshold can lower security, but provides more flexibility in the system. The user can then select an appropriate number of witnesses based on the expected density of their network, and use the depth parameter to find an appropriate trade-off between security and reliability, and communication overhead and flexibility.

\begin{figure*}[h]
     \centering
     \begin{subfigure}[t]{0.7\textwidth}
            \centering
            \includegraphics[width=\textwidth]{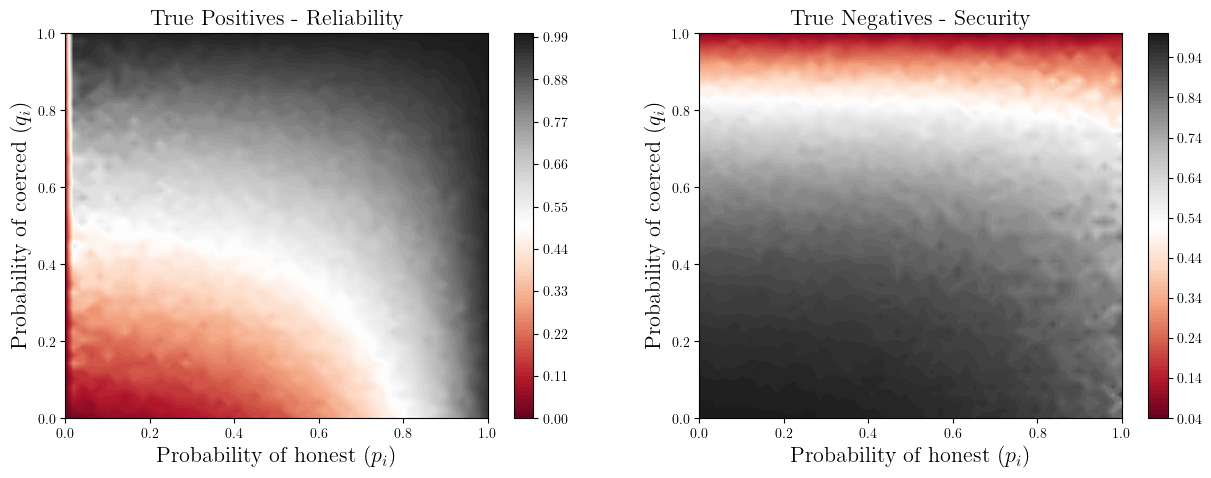}
            
     \end{subfigure}
     \hfill
     \begin{subfigure}[t]{0.7\textwidth}
             \centering
            \includegraphics[width=\textwidth]{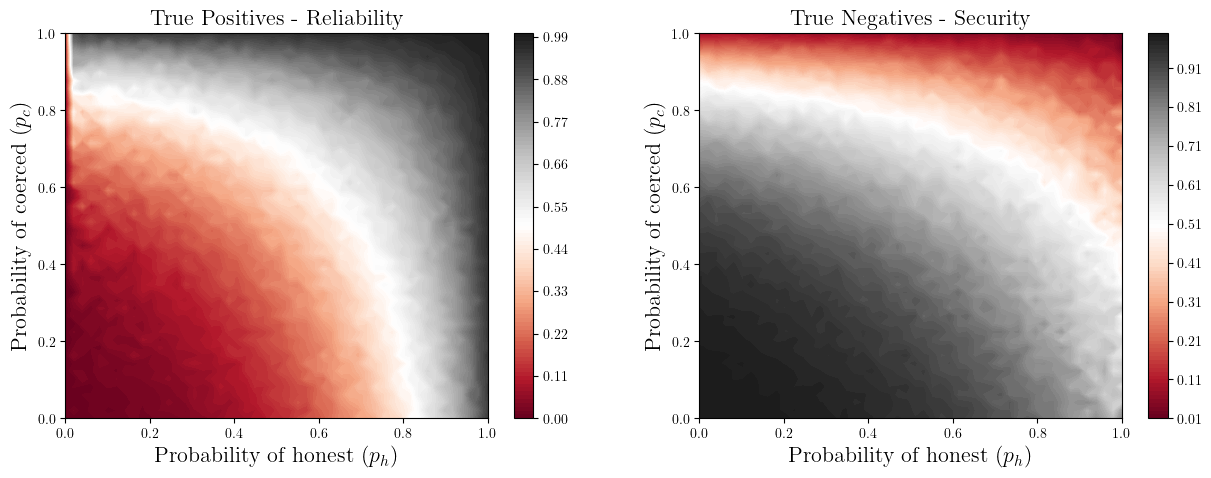}
         
     \end{subfigure}
     
        \caption{T-PoP performance for {\bf agent-based model}. The panels on the left show reliability $R$, while the panels on the right show security, $S$. The first row is associated with model parameters $\theta_1$, the second row is associated with model parameters $\theta_2$. Notice the close similarity to Figure \ref{tpop model figure}.}
    \label{fig: tpop simulation}
\end{figure*}

\begin{figure*}[h]
     \centering
     \begin{subfigure}[t]{0.7\textwidth}
            \centering
    \includegraphics[width=\textwidth]{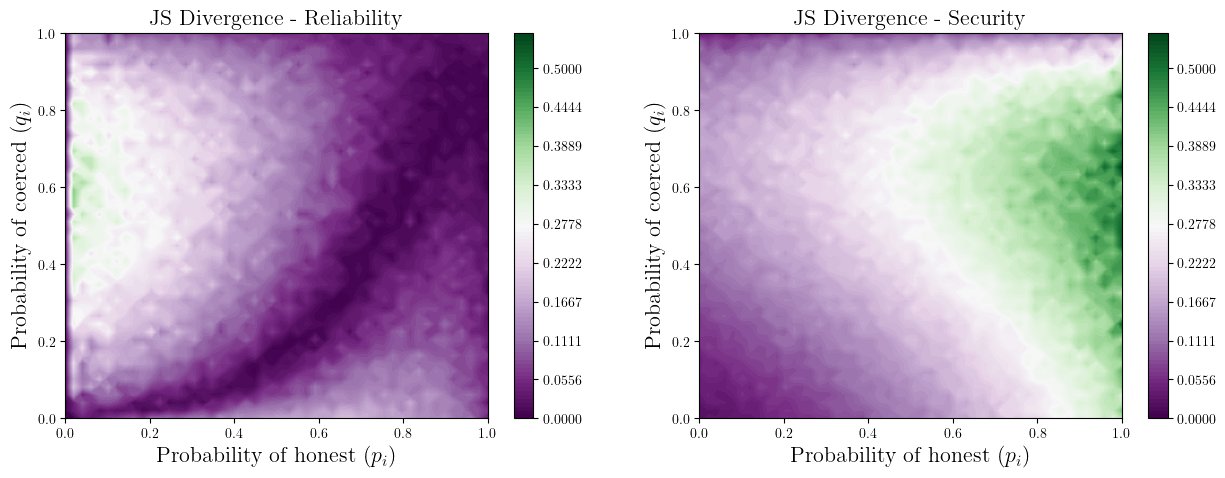}

     \end{subfigure}
     \hfill
     \begin{subfigure}[t]{0.7\textwidth}
             \centering
    \includegraphics[width=\textwidth]{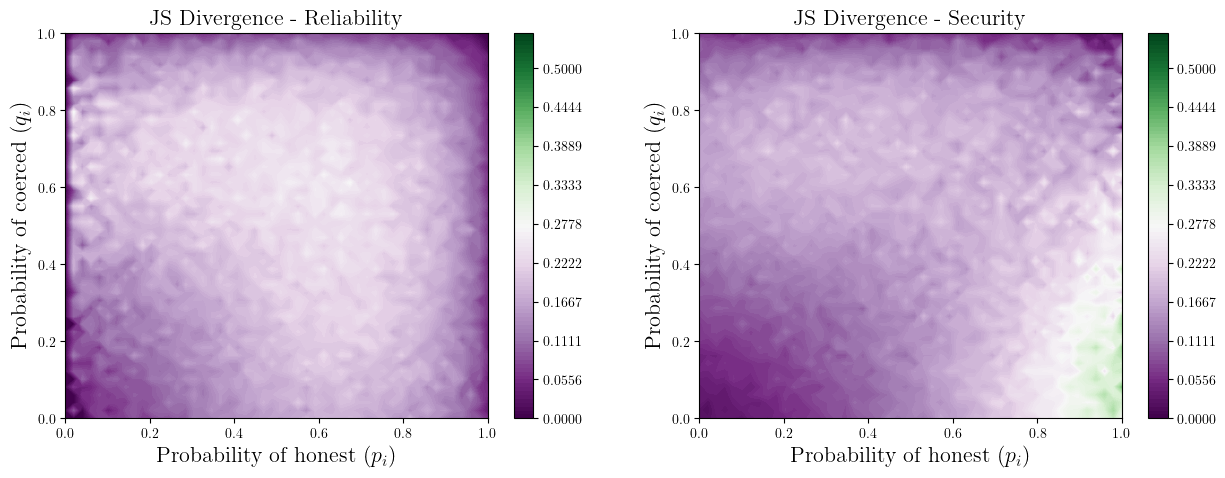}

     \end{subfigure}
     
        \caption{Jensen-Shannon divergence (JSD) between $R_s$ and $R_m$ (left column) and between $S_s$ and $S_m$ (right column) for $\theta_1$ (top row) and $\theta_2$ (bottom row).}
    \label{fig:tpop verification}
\end{figure*}

\subsection{Preliminary results}
Our objective in this section is twofold. On the one-hand, we want to show some preliminary results on the performance of T-PoP for a given choice of operating conditions. On the other hand,  we are interested in validating the results from the probabilistic graphical model presented in the previous section, with a view to creating an analytical framework  for analysis of the T-PoP class of algorithms. This gives us confidence that the results obtained for simple model parameter settings (e.g.\ $d$ small) still hold in more realistic scenarios.

The simulations have been set up as follows: we considered each possible combination of $p_h$ and $p_c$ in the ranges $[0,1]$, with steps of 0.02. For each combination we ran 50 Monte Carlo simulations and we computed empirical estimates of the values of $R(p_h, p_c)$ and $S(p_h, p_c)$. Simulations are set up in such a way that on average each agent has 50 neighbours in their range of sight $r_i$. While this number might appear very high, we wanted to make sure that the results obtained were comparable to the ones obtained with the probabilistic graph model. Moreover, real-life situations with high density of pedestrians (e.g., the underground during peak hours) would map well into this scenario. We ran these simulations for the choice of parameters  $\theta_1$ and $\theta_2$.

The results are shown in Figure \ref{fig: tpop simulation}. While T-PoP with $\theta_1$ yields better performance overall (as both $R(p_h, p_c)$ and $S(p_h, p_c)$ are higher for each choice of $p_h$ and $p_c$) the second set of simulations shows that decreasing the number of witnesses by a third and increasing the depth level by 1 allows us to achieve similar results. This is useful because---while the total number of nodes in each prover's tree is the same for both scenarios---a tree of depth 2 with 2 witnesses per parent places  a smaller communication overhead on the prover, because it only needs to name 2 witnesses, as opposed to 6. In this way, the load is shared among the prover and the witnesses.

Overall, in high density scenarios, the results of both simulations show that---if $p_h > 0.9$ and $p_c < 0.2$---T-PoP is capable of achieving $S> 0.85\%$ and $R> 0.9 \%$ for $\theta_1$, and $S> 0.7\%$ and $R > 0.9\%$ for $\theta_2$.

For lower proportions of honest agents and higher proportions of coerced agents (i.e.\ in the presence of many colluding, dishonest and coerced agents), the performance of T-PoP degrades. This is to be expected in a decentralised system such as T-PoP, since it is virtually impossible to distinguish between a group of honest agents verifying each other and a group of dishonest and coerced agents collaborating to verify each other in a fraudulent manner. 
Accordingly, we can observe across all figures that---even when the percentage of honest agents is low---the security remains high at the expense of reliability. We observe that---whilst, indeed, T-PoP can detect true negatives (i.e.\ be secure) in highly (and perhaps even unrealistically highly) adversarial environments---the drawback is that it penalises honest agents too harshly (i.e.\ is unreliable). This is a consequence of the collaborative nature of the algorithm. When the number of honest agents in the system is low (i.e.\ $p_h \downarrow 0$), they will---with high probability (w.h.p.)---be misclassified as dishonest because they  will select dishonest witnesses w.h.p. 

\subsection{Validation of the graphical model (Figure~\ref{fig:Probability Model})}

For validation of the graphical probability model, we make use of the Jensen-Shannon Divergence (JSD) \cite{menendez1997jensen} to  quantify the distance between the probability distributions obtained through the agent-based model (i.e.\ the T-PoP implementation) and the graphical  model. In what follows, we refer to the values of $R$ and $S$ obtained from the simulated agent-based model as $R_s$ and $S_s$, and the ones obtained from the graphical model as $R_m$ and $S_m$. We compute two JSD-based metrics: (i)  the $(p_h,p_c)$-indexed (i.e.\ pointwise) JSD map between $R_m$ and $R_s$, and between $S_m$ and $S_s$, respectively; and (ii) the global JSD between the normalized $R_m$ and $R_s$ maps, and the normalized $S_m$ and $S_s$ maps, respectively. By ``normalized'', we mean that each of these positive maps is divided by its element sum, yielding a probability mass function (pmf). In case (ii),  we can therefore condense into a single number the difference between the performance figures ($R$ and $S$, respectively) for the simulated T-PoP system and its graphical model (Figure~\ref{fig:Probability Model}).

The results for the point-wise evaluation ((i) above) of the JSD are shown in Figure \ref{fig:tpop verification}, while the global evaluation ((ii) above) is summarised in table \ref{tab:JSD table}. Note that $0\leq$ JSD $\leq 1$, with lower values achieved when probabilities are close in value (i.e.\ in cases of good agreement between the behaviour of the simulated system and the graphical model). It is clear that---at least for high density scenarios---the behaviour of the graphical model closely mirrors that of the implemented T-PoP system. Nevertheless, the pointwise JSD results reveal significant discrepancies in the security ($S$) metric when $p_h \uparrow 1$(i.e.\ for high proportions of honest users).  

\begin{table}[h]
    \centering
    \def\arraystretch{1.5}
    \begin{tabular}{|c|c|c|}
        \hline
        Parameters & $JSD(R_m, R_s)$ & $JSD(S_m, S_s)$ \\
        \hline
        $\theta_1$  & 0.139 & 0.095\\
        \hline
        $\theta_2$ & 0.174 & 0.059\\
        \hline
 
        \hline
         
    \end{tabular} 
    \caption{Jensen-Shannon divergence (JSD) between the normalized reliability ($R$) and security ($S$) maps for two sets of model parameters.}
    \label{tab:JSD table}
\end{table}

\section{Conclusion}
We have presented a proof-of-position class of algorithms that are fully decentralised. They can be run by any agent participating in the network and they do not assume trust in a central authority, nor do they rely on physical infrastructure. We also considered a range of attack vectors by allowing agents not only to lie about their own position, but also about others' positions. Our algorithm can also be computed in a privacy-preserving manner, as there is no need for the true location of an agent to be revealed to the network. We also developed a theoretical graphical model for this class of proof-of-position algorithms, and statistically validated  the model via comparative analysis of their respective performances.
In future work, we will use the theoretical model to predict the performance of T-PoP as a function of its operating conditions, $\theta$. Specifically,  we will be interested in characterising the effect of the depth ($d$), threshold ($t$) and number of witnesses ($w_l$) on the security and reliability of the T-PoP class of algorithms. Developing such a framework can allow users to select the optimal operating conditions of the algorithm to meet their needs, based on their expected density, fault tolerance and proportion of honest and non-coerced agents  in their system. The theoretical model will also allow performance guarantees to be deduced for T-PoP. Finally, we intend   to explore the suitability of T-PoP for specific use-cases in the presence of more complex adversarial scenarios. 

\textbf{Acknowledgments:} The authors would like to thank the IOTA Foundation for funding this research. 

\bibliographystyle{ieeetr}
\bibliography{mybib}

\end{document}